\documentclass[prb,twocolumn,showpacs,superscriptaddress,amsfonts]{revtex4}

\usepackage{psfig}

\renewcommand{\vec}[1]{{\rm\bf #1}}
\renewcommand{\cite}[1]{[\onlinecite{#1}]}
\newcommand{\ep}{\epsilon}
\newcommand{\vep}{\varepsilon}
\newcommand{\hc}{\mathrm{h.\;c.}}

\begin{document}

\title{Coulomb blockade in quantum dots under AC pumping}
\author{D.~M.~Basko}
\email{basko@ictp.trieste.it}
\affiliation{The Abdus Salam International Centre for Theoretical Physics,
Strada Costiera 11, 34100 Trieste, Italy}
\author{V.~E.~Kravtsov}
\affiliation{The Abdus Salam International Centre for Theoretical Physics,
Strada Costiera 11, 34100 Trieste, Italy}
\affiliation{Landau Institute for Theoretical Physics,
2 Kosygina Street, 117940 Moscow, Russia}

\date{\today}
\begin{abstract}
We study conductance through a quantum dot under Coulomb blockade
conditions in the presence of an external periodic perturbation.
The stationary state is determined by the balance between the
heating of the dot electrons by the perturbation and cooling.
We analyze two cooling mechanisms: electron exchange with the
cold contacts and emission of phonons. Together with the usual
linear Ohmic heating of the dot electrons we consider possible
effects of dynamic localization. The combination of the
abovementioned factors may result in a drastic change of the
shape of the Coulomb blockade peak with respect to the usual
equilibrium one.
\end{abstract}

\pacs{73.21.La, 73.23.-b, 73.20.Fz, 78.67.Hc}

\maketitle

\section{Introduction}

At low temperatures electronic conduction through a quantum dot
weakly coupled to the contacts is governed by the Coulomb blockade
effect~\cite{MarcusRev} -- suppression of transport due to the
energy cost of changing the number of electrons in the dot.
Efficient conduction through such a dot is possible only when the
electrostatic potential of the dot, controlled by external gates,
is tuned to a special value where the Coulomb energies of the
states with $N$~and~$N+1$ electrons in the dot are close for
some~$N$. As a result, the linear response conductance exhibits a
sharp peak as a function of the gate voltages. Theory of the
Coulomb blockade in equilibrium is well developed by
now~\cite{Aleinerrev}.

In the last few years several experiments have been done on
quantum dots under an external ac perturbation~\cite{Marcus}.
Under these non-equilibrium conditions the electronic temperature
of the dot is no longer determined by the external cryostat, but
by the balance between heating by the ac perturbation and cooling
due to various mechanisms. At sufficiently low temperatures
cooling is dominated by simple electronic exchange between the
dot and the contacts (the latter are assumed to be maintained at
a constant low temperature determined by the cryostat).
In this case, as the gate voltage is tuned away from the Coulomb
blockade peak, the cooling rate changes, so does the electronic
temperature, thus changing the peak shape with respect to the
equilibrium one.
This simple qualitative consideration poses the problem, which is
going to be studied in detail in the present work.

Another motivation to study these effects is the search for
experimental signatures of dynamic localization~(DL).
Experimental observation of DL in trapped ultracold atoms in
the field of a modulated laser standing wave~\cite{Raizen}
provided a solid ground for the preceding extensive theoretical
studies of the kicked quantum rotor~\cite{Izrailev,Haake}.
In a recent publication~\cite{us} we have shown that an analogous
suppression of the energy absorption is possible for a solid-state
system -- a chaotic quantum dot under an ac excitation, e.~g. like
those used in experiments of Ref.~\cite{Marcus}, which makes the
question about the possibility of observation of DL in a quantum
dot highly relevant. If one wishes to detect this effect by
transport measurements, the Coulomb blockade regime is the most
suitable, since it is in this regime that the transport is
sensitive to the internal state of the dot, while for an open
dot, when electron-electron interaction can be neglected, the
conductance is insensitive to the electron energy distribution
in the dot~\cite{Vavilov,Kanzieper}.

First, consider the standard picture of heating by an ac
perturbation.
Let the single-electron mean level spacing~$\delta$ in the dot
be small enough. Then, if an external time-dependent periodic
perturbation with the frequency~$\omega$ is applied, the total
electronic energy~$E$ in the dot (counted from that of the ground
state) grows linearly with time as described by the Fermi Golden
Rule: $E(t)=\Gamma\omega^2t/\delta\equiv{W}_0t$. The
probability of each single-electron transition per unit time,
denoted by $\Gamma$, measures the strength of the
perturbation~\cite{perturbation}. The criterion of validity of
the Fermi Golden Rule is $\delta\ll\Gamma$, and $\Gamma\ll\omega$
is also assumed ($\hbar=1$). This picture corresponds also to the
classical Ohmic absorption by a small particle made of a metal
with large conductivity $\sigma\gg\omega$.

This picture (hereafter referred to as Ohmic absorption) is valid
provided that each act of photon absorption by an electron is
independent of the previous ones; however, for a discrete energy
spectrum this turns out not to be the case. After many transitions
the absorption rate decreases due to accumulation of the quantum
interference correction~\cite{us}, so that after a time
$t_*\sim\Gamma/\delta^2$ the absorption is completely suppressed.
This effect was named the dynamic localization in energy
space; the effective electronic temperature (the characteristic
spread of the electron distribution function), reached by the
time~$t_*$, $T_*\sim\Gamma\omega/\delta$, plays the role of the
localization length. Note that DL has nothing to do with the
saturation of absorption by a pumped two-level system, as in our
case the spectrum is unbounded. DL is the consequence of level
discreteness: at $\delta\rightarrow{0}$ it takes longer time for
the DL to develop, and for the continuous spectrum there is no~DL.
Since this effect drastically modifies the heating rate, the
stationary state of the dot is strongly affected.

The considerations of Ref.~\cite{us} were based on random matrix
theory description of the single-particle properties of the dot.
This description is valid provided that all energy scales in the
problem are small compared to the Thouless energy~$E_{Th}$
(defined by the order of magnitude as the inverse of the time
required for an electron to travel across the dot and thus to
randomize its motion due to scattering off the dot boundaries).
For the dot to be in the Coulomb blockade regime, the effective
temperature should be also smaller than the dot Coulomb charging
energy~$E_c$. Thus, in the following, the hierarchy of scales
$\delta\ll\Gamma\ll\omega\ll{T}_*\ll{E}_{Th},E_c$ is assumed.
Note that in the random matrix theory one can neglect multiphoton
processes as they are of the order of the inverse matrix size.

Possible cooling mechanisms for electrons in the dot are
(i)~electron exchange with the contacts, and (ii)~energy exchange
with the phonon subsystem. Both electrons in the contacts and
phonons in the dot are assumed to be maintained at a constant
temperature~$T_0$ determined by the cryostat.
In the following we analyze the interplay of the abovementioned
effects in heating and cooling, and see how they affect the shape
of the Coulomb blockade peak.
In a short preliminary version of this study we have considered
only the first cooling mechanism~\cite{BK}. Here we include
cooling by phonon emission which, to the best of our knowledge,
has been little studied for a quantum dot.

The paper is organized as follows. In Sec.~II we analyze the
heating and discuss how it is affected by dynamic localization.
Sections III~and~IV are
dedicated to a detailed analysis of the two cooling mechanisms.
In Sec.~V we consider the resulting stationary state and the
Coulomb blockade peak shape. Finally, in Sec.~VI we summarize
the main results.

\section{Heating by ac perturbation}\label{DL}

In the Ohmic regime the energy absorption by electrons is
linear in the field intensity and given by
$W_0=\Gamma\omega^2/\delta$ (we remind the reader that
$\Gamma$~is a measure of the microwave field intensity,
equal to the probability per unit time of a single one-photon
transition). The same expression can be obtained from simple
classical arguments considering a small particle made of a
metal with a large finite conductivity $\sigma\gg\omega$.

In the regime of the strong dynamic localization the absorption
is no longer given by the simple Ohmic expression. For
non-interacting electrons in a closed dot the absorption
becomes completely suppressed by interference corrections
that develop in a characteristic time $t_*\sim\Gamma/\delta^2$,
and the effective temperature of the electrons, reached by that
time, is $T_*\sim\Gamma\omega/\delta$. In the presence of
{\em weak} dephasing processes with the dephasing
rate~$\gamma_{\phi}\ll{1/t_*}$
there is a residual absorption with the rate given by
\begin{equation}\label{WinDL=}
W_{\rm{in}}\sim{W}_0\gamma_{\phi}t_*=
T_*^2\,\frac{\gamma_{\phi}}{\delta}\:.
\end{equation}
If the dephasing is too strong, $\gamma_{\phi}\gtrsim{1}/t_*$,
the dynamic localization is destroyed and $W_{\rm{in}}\sim{W}_0$.

The expression~(\ref{WinDL=}) was justified in Ref.~\cite{Basko}
for the dephasing due to electron-electron collisions.
The main condition of its applicability is that dephasing should
be a sequence of distinct phase-destroying events with average
frequency~$\gamma_{\phi}$, rather than phase diffusion, in
which case the dephasing rate roughly coincides with the
quasiparticle relaxation rate: $\gamma_{\phi}\sim\gamma_{qp}$.
This is certainly correct for the case of electron escape to
the contacts, since in this case the electron is effectively
replaced by another one with an absolutely random phase. This is
also true for electron-electron and electron-phonon collisions,
since the typical energy transfer during a collision is of the
order of the (effective) electronic temperature in the dot,
which is large: $T\gg{1/t_*},\gamma_{qp}$ (this inequality
follows from $T_*\gg{1}/t_*$ due to $\Gamma,\omega\gg\delta$,
and from $T\gtrsim{T}_*$).

Once the condition $\gamma_{\phi}\sim\gamma_{qp}$ is verified,
the following consideration can be applied. As the collisions
are rare ($\gamma_{qp}t_*\ll{1}$), the electrons spend most of
the time in the states localized in energy space, having
definite phase relationships. When at some moment the phase of
some electron is destroyed, its wave packet starts spreading
along the energy axis. It localizes again after the time
$\sim{t}_*$, in the meantime spreading by $\sim{T}_*$.
Thus, the ac driven dynamics following the collision leads to
a change of the total electronic energy of $\sim{T}_*$~per
collision. The sign of this change is, however, arbitrary,
because a periodic perturbation can equally cause transitions
up and down the spectrum. Only the presence of the filled Fermi
sea below (i.~e., an energy gradient of the electronic
distribution function) makes absorption the preferred direction,
which means that if the electronic temperature $T\gg{T}_*$, the
energy absorbed per collision is on the average $\sim{T}_*^2/T$
rather than~$T_*$. The effective number of electrons that can
participate in a collision is $\sim{T}/\delta$ (due to the
degenerate Fermi statistics).
During the time interval $\sim{1}/\gamma_{qp}$ each of these
electrons participates in one collision, so the total number
of collisions per unit time is $\sim(T/\delta)\gamma_{qp}$.
This gives the energy absorption rate
$W_{\rm{in}}\sim({T}_*^2/T)(T/\delta)\gamma_{qp}$, which is
exactly Eq.~(\ref{WinDL=}).

\begin{figure}
\psfig{figure=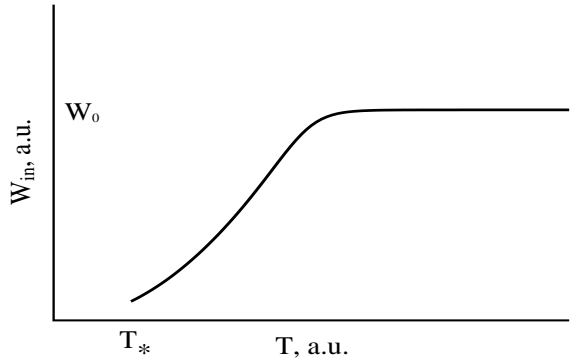,width=9cm,height=6cm}
\caption{\label{Win:}
A schematic view of the dependence of the absorption rate
$W_{\rm{in}}$ on the effective electronic temperature in the
dot: in the dynamic localization regime the absorption is
due to dephasing, so it is temperature-dependent; when the
temperature becomes high enough, the dephasing destroys the
dynamic localization and the absorption is Ohmic.}
\end{figure}

The same can be seen from an alternative argument.
After each collision the electron spends the time $\sim{t}_*$
absorbing the energy from the microwave field, then it stops to
absorb (the dynamic localization occurs) and waits for the next
event (provided that $t_*\ll{1}/\gamma_{\phi}$). Thus the
absorption rate of the whole system is given by the simple
weighted average: $W_{\rm{in}}\sim{W}_0\gamma_{\phi}t_*$, which
is again Eq.~(\ref{WinDL=}).

An important point is that dephasing rate, generally speaking,
depends on the electronic temperature, which results in a
temperature-dependent absorption rate in the DL regime
(Fig.~\ref{Win:}).
The temperature, in turn, determined by the balance between
energy absorption and cooling.
This feedback leads to a non-trivial dependence of the
characteristics of the stationary state on the control parameters,
which will manifest itself in a change of the Coulomb blockade
peak shape, as will be shown below. The absorption itself becomes
nonlinear with the field intensity through the dependence of
$\gamma_{\phi}$~on~$W_0$.

\section{Cooling due to electron escape}

We characterize the coupling of the dot to the two contacts by
single-particle escape rates $\gamma_1$~and~$\gamma_2$.
When they are much smaller than the mean single-particle level
spacing~$\delta$ in the dot, the fluctuations of the total
charge on the dot are small.
If the dot is coupled to several gates through
capacitances~$C_i$ and voltages~$V_i$ are applied to the gates,
the electrostatic energy of the dot with $N$~electrons on it
is given by
\begin{equation}
E(N)=\frac{e^2N^2}{2C}+\sum_i\frac{C_iV_i}{C}\,eN\:,\quad
C\equiv\sum_iC_i\:,
\end{equation}
where $e^2/(2C)\equiv{E}_c$ is the charging energy.
The energy cost of adding an electron is
\begin{equation}
U\equiv{E}(N+1)-E(N)=\frac{e^2}{C}
\left[N+\frac{1}{2}+\sum_i\frac{C_iV_i}{e}\right].
\end{equation}
If all gates have the same voltage, then (up to a constant)
$U$~is given by this voltage. Generally, we will call~$U$
{\em the reduced gate voltage}; it is a natural control
parameter for the system.
If the expression in the brackets is of the order of unity
and the temperature $T\ll{E}_c$, the conductance through
the dot is suppressed due to the Coulomb blockade.
If the gate voltages are tuned so that the expression in the
brackets is small for some particular~$N$, the dot
conductance~$G(U)$ exibits a peak for these values of~$U$.
The width of the peak $\Delta{U}\sim{T}$, which can be used
to measure the temperature of the system.

\subsection{Sequential tunneling}

When $U$~is tuned to the peak, the main contribution to the
conductance comes from the leading order of the perturbation
theory in the dot-contact coupling. For characteristic
temperatures $T\gg\delta$ one can describe the system by
rate equations of Kulik and Shekhter~\cite{Shekhter}. We
consider these equations for the case when the electron
energy distribution function in the dot~$f_{\ep}$ is
non-equilibrium. Let the distribution in the $\alpha$th
contact be~$f^{(\alpha)}_{\ep}$. Assuming the dot to
have either~$N$ or $N+1$ electrons with the probabilities
$p_N$,~$p_{N+1}$ to have $N$~or $N+1$
electrons on the dot (all others are neglected, so
$p_N+p_{N+1}=1$), we can write the rate equation as
\begin{eqnarray}\nonumber
\frac{dp_N}{dt}&=&2p_{N+1}\sum_{\alpha=1,2}\gamma_{\alpha}
\int{f}_{\ep}(1-f^{(\alpha)}_{\ep+U})\,\frac{d\ep}{\delta}-\\
&&-2p_N\sum_{\alpha=1,2}\gamma_{\alpha}
\int(1-{f}_{\ep})f^{(\alpha)}_{\ep+U}\,\frac{d\ep}{\delta}\:,
\end{eqnarray}
where the factor of two comes from the spin degeneracy.
The distributions in the contacts are assumed to be Fermi-Dirac
ones with the temperature~$T_0$:
\begin{equation}\label{FermiDirac=}
f^{(\alpha)}_{\ep}=f^{T_0}_{\ep}\equiv\frac{1}{e^{\ep/T_0}+1}\:.
\end{equation}
As usual, we require $p_N$~and~$p_{N+1}$ to be stationary.
Shifting the distribution in one of the contacts by an
infinitesimal voltage, one obtains the linear response
conductance~$G$:
\begin{widetext}
\begin{eqnarray}\label{conductance=}
&\displaystyle G(U)=\frac{2e^2}{\delta}\,
\frac{\gamma_1\gamma_2}{\gamma_1+\gamma_2}\,
\frac{F_{\rm{in}}^2(U)\,F_{\rm{out}}'(U)
-F_{\rm{in}}'(U)\,F_{\rm{out}}^2(U)
+F_{\rm{in}}(U)\,F_{\rm{out}}(U)}
{[F_{\rm{in}}(U)+F_{\rm{out}}(U)]^2}\:,&\\
&\displaystyle
F_{\rm{in}}(U)\equiv
\int(1-f_{\ep})f^{T_0}_{\ep+U}\,d\ep\,,\quad
F_{\rm{out}}(U)\equiv
\int{f}_{\ep}(1-f^{T_0}_{\ep+U})\,d\ep\,.&
\end{eqnarray}
\end{widetext}
In the equilibrium case, when $f_{\ep}=f_{\ep}^{T_0}$ as well,
the last fraction in the right-hand side of
Eq.~(\ref{conductance=}) reduces to the familiar expression
$(1/2)(U/T)/\sinh(U/T)$.

Tunneling events lead to the change in the distribution function
in the dot. The kinetic equation describing this process can
be obtained straightforwardly from equations of Ref.~\cite{Shekhter}
and reads as
\begin{eqnarray}\nonumber
&&\frac{\partial{f}_{\ep}}{\partial{t}}=(\gamma_1+\gamma_2)\times\\
\label{kinseq=}&&\times
\frac{(1-f_{\ep})f^{T_0}_{\ep+U}\,F_{\rm{out}}(U)
-f_{\ep}(1-f^{T_0}_{\ep+U})\,F_{\rm{in}}(U)}
{F_{\rm{in}}(U)+F_{\rm{out}}(U)}\:.\quad
\end{eqnarray}
If we introduce the functions
\begin{eqnarray}
\mathcal{E}_{\rm{in}}(U)\equiv
\int(1-f_{\ep})f^{T_0}_{\ep+U}\,\ep\,d\ep\,,\\
\mathcal{E}_{\rm{out}}(U)\equiv
\int{f}_{\ep}(1-f^{T_0}_{\ep+U})\,\ep\,d\ep\,,
\end{eqnarray}
and denote by $\gamma\equiv\gamma_1+\gamma_2$ the total
single-electron broadening, the cooling rate for the dot
electrons (the total energy loss per unit time) can be written as:
\begin{equation}\label{cooling=}
W_{\rm{out}}(U)=\frac{\gamma}{\delta}\,
\frac{\mathcal{E}_{\rm{out}}(U)\,F_{\rm{in}}(U)-
\mathcal{E}_{\rm{in}}(U)\,F_{\rm{out}}(U)}
{F_{\rm{in}}(U)+F_{\rm{out}}(U)}\:.
\end{equation}
From the kinetic equation~(\ref{kinseq=}) one can also extract the
single-particle escape rate for a particle with the energy~$\ep$:
\begin{equation}
\gamma_{esc}=\gamma(1-f^{T_0}_{\ep+U})\,\frac{F_{\rm{in}}(U)}
{F_{\rm{in}}(U)+F_{\rm{out}}(U)}\:.
\end{equation}
In the following we will use the expression for $\gamma_{esc}$
at $\ep=0$ as an estimate.
We will also use the Fermi-Dirac form for the electronic
distribution function with some temperature~$T$. This is true
only if electron-electron collisions restore the Fermi-Dirac
shape much faster than it is modified by other processes. If
this is not the case, $T$~still gives the characteristic width
of the distribution function.
It is determined by the balance between heating by the ac field
and cooling considered in the previous section.

We also assume the electronic temperature in the dot to be much
higher than the temperature of the contacts (the latter can be
made as low as $\sim{10}$~mK~\cite{Kouwenhoven2003}), which is
true if the pumping power is high enough. Then we can set the
temperature of the contacts to be zero, which allows an explicit
calculation in Eqs.~(\ref{conductance=})--(\ref{cooling=})
[we denote $x\equiv{U}/(2T)$, $G_0\equiv{G}(U=0)$]:
\begin{eqnarray}
&&\displaystyle
F_{\rm{in}}(U)=T\ln\left[1+e^{-2x}\right],\\
&&F_{\rm{out}}(U)=F_{\rm{in}}(-U),\\
\label{GVG0=}
&&\displaystyle\frac{G(U)}{G_0}=1-\frac{x\tanh{x}}{\ln(2\cosh{x})}\:,\\
&&\mathcal{E}_{\rm{in}}(U)=-2T^2\int\limits_{x}^{\infty}
(1-\tanh{y})y\,dy\:,\\
&&\mathcal{E}_{\rm{out}}(U)=-\mathcal{E}_{\rm{in}}(-U),\\
\nonumber
&&\frac{W_{\rm{out}}(U)}{(\gamma/\delta)T^2}=\frac{\pi^2}{12}-x^2+\\
\label{WoutCBP=}
&&+\frac{2x}{\ln(2\cosh{x})}\int\limits_0^x{y}\tanh{y}\,dy\:,\\
\label{gescCBP=}
&&\frac{\gamma_{esc}(U)}{\gamma}=
\frac{1}{2}-\frac{|x|}{2\ln(2\cosh{x})}\:.
\end{eqnarray}

\subsection{Inelastic cotunneling}

At large~$U\gg{T}$ sequential tunneling becomes suppressed
exponentially.
In this situation both the conduction and cooling become
dominated by cotunneling -- a second-order process whose
probability contains an additional small factor
$\gamma\delta/U^2$. Obviously, only inelastic
cotunneling~\cite{Averin} can contribute to cooling.
Elastic cotunneling~\cite{GlazmanMatveev}, which does not
change the electronic state of the dot, contributes to
conduction at temperatures $T<\sqrt{{E}_c\delta}$. We
will be interested in higher temperatures and do not
consider this contribution.

A straightforward generalization of the considerations of
Ref.~\cite{Averin} to the non-equilibrium case leads to
the following expression for the conductance in terms of
the electronic distribution functions and the kinetic
equation for the distribution in the dot:
\begin{widetext}
\begin{eqnarray}
&&G(U)=\frac{4e^2\gamma_1\gamma_2}{\pi{U}^2\delta^2}
\int{f}_{\ep-\Omega}(1-f_{\ep})(1-f^{T_0}_{\ep'})
\left(-\frac{\partial{f}^{T_0}_{\ep'+\Omega}}{\partial\ep'}
\right)d\ep\,d\ep'\,d\Omega\,,\\
\label{kincot=} &&\frac{\partial{f}_{\ep}}{\partial{t}}=
\frac{\gamma^2}{\pi{U}^2\delta}\int\left[
(1-f_{\ep})f_{\ep-\Omega}(1-f^{T_0}_{\ep'})
f^{T_0}_{\ep'+\Omega}
-f_{\ep}(1-f_{\ep-\Omega})f^{T_0}_{\ep'}
(1-f^{T_0}_{\ep'+\Omega})\right]d\ep'\,d\Omega
\end{eqnarray}
\end{widetext}
For a Fermi-Dirac distribution, $f_{\ep}=f^T_{\ep}$ the integrals
can be calculated explicitly for any temperatures $T,T_0$:
\begin{eqnarray}
&&G(U)=\frac{2\pi{e}^2\gamma_1\gamma_2}{3{U}^2\delta^2}\,(T^2+T_0^2)\,,\\
&&W_{\rm{out}}(U)=\frac{2(\gamma_1+\gamma_2)^2}{15\pi\delta^2U^2}\,
(T^4-T_0^4)\,.
\end{eqnarray}
The electron escape rate at $\ep=0$ can be extracted from the
kinetic equation~(\ref{kincot=}):
\begin{equation}
\gamma_{esc}=\frac{\pi}{6}\,\frac{\gamma^2}{U^2\delta}\,(T^2+2T_0^2)\,.
\end{equation}
If we set, as before, $T_0=0$, we obtain
the following explicit expressions:
\begin{equation}\label{WoutCBV=}
\frac{W_{\rm{out}}}{(\gamma/\delta)T^2}=
\frac{\pi^3}{30}\frac{\gamma/\delta}{x^2}\,,\quad
\frac{G}{G_0}=\frac{\pi}{6}\,\frac{\gamma/\delta}{x^2}\,,\quad
\frac{\gamma_{esc}}{\gamma}=\frac{\pi}{24}\,\frac{\gamma/\delta}{x^2}\,.
\end{equation}

\subsection{Photon-assisted tunneling}

So far the only effect of the AC~perturbation we were interested
in was to cause transitions between single-particle states in the
dot. The perturbation, however, may possess a component
$V\cos\omega{t}$, proportional to the unit matrix in the dot
single-particle Hilbert space. In a closed dot this component does
not cause any transitions and can be gauged out completely, so it
does not affect any observables, either single-particle or
many-particle ones (in particular, it does not affect
electron-electron collisions).

However, when the dot is connected to contacts, this is no longer
the case, as the diagonal component is responsible for the
photon-assisted tunneling~\cite{TienGordon}. This effect can be
taken into account by replacing the dot electron distribution
function~$f_{\ep}$ in the above formulas by
\begin{equation}\label{PAT=}
f_{\ep}\rightarrow
\sum_{n=-\infty}^{\infty}J_n^2(V/\omega)\,f_{\ep+n\omega},
\end{equation}
where $J_n$~is the Bessel function. Photon-assisted tunneling
will not be important for our considerations if the smearing of
the distribution function given by Eq.~(\ref{PAT=}) is much
smaller than the thermal smearing. Using the asymptotic
expansion of $J_n(z)$ at large~$n$, this condition can be
written as
\begin{equation}
J_n(z)\sim\frac{1}{\sqrt{2\pi{n}}}\left(\frac{ez}{2n}\right)^n
\;\;\;\Rightarrow\;\;\;\max\{V,\omega\}\ll{T}\,.
\end{equation}
The condition $\omega\ll T$ is authomatically fulfilled if 
$\Gamma\gg\delta$ and $T>T_{*}=\Gamma\omega/\delta$. As for the
condition $V\ll T$ we note that within the $N\times N$ random
matrix  approximation, adopted in Ref.~\cite{BK}, we have 
$\langle V^{2}\rangle=(1/N)\Gamma\delta$ so that $V\rightarrow{0}$
as $N\rightarrow\infty$.

Besides the random component with zero mean included in the
random-matrix treatment, $V$~can have a deterministic part. It is
given by the spatial average of the perturbation potential over
the dot volume, and enters our model as an {\em independent}
parameter. Thus in order to fulfill the condition~$V\ll{T}$ a
special experimental care should be taken.

\section{Cooling due to phonon emission}

\subsection{General expressions}\label{PhononsGeneral}

Another important mechanism of electronic energy relaxation
is emission of phonons. For mesoscopic metallic rings with
diffusive electronic motion this problem was addressed in
Ref.~\cite{Yudson}.
For quantum dots energy relaxation at frequencies smaller
than the mean level spacing has been
considered~\cite{Altshuler,Galperin}; here we are interested
in the opposite limiting case, $\delta$~being the smallest
energy scale.
Below we estimate the corresponding cooling rate for clean
(ballistic) quantum dots made out of 2D~electron gas (2DEG)
in a GaAs/AlGaAs heterostructure~\cite{Marcus} and bulk
3D~phonons.

For ballistic dots (whose size~$L$ is smaller than the
elastic mean free path~$\ell$) one does not need to take
into account phonon-induced impurity
displacements~\cite{Reizer}, so the phonon-induced potential
felt by the electrons can be written in the form
\begin{equation}
\hat{V}(\vec{r})=\int\frac{d^3\vec{q}}{(2\pi)^3}\,
\hat{V}(\vec{q})\,e^{i\vec{q}\vec{r}}=
\sum_{\vec{q},\lambda}V_{\vec{q},\lambda}
\hat{b}_{\vec{q},\lambda}e^{i\vec{q}\vec{r}}+\hc,
\end{equation}
where 
$\hat{b}_{\vec{q},\lambda}$~is the annihilation operator
for a phonon mode~$\lambda$ with the wave vector~$\vec{q}$.
The detailed form of the coupling $V_{\vec{q},\lambda}$
depends on the specific coupling mechanism to be specified
below.

The probability of the electronic transition from an initial
single-particle state~$s$ with the energy~$\ep_s$ and the
wave function~$\psi_s(\vec{r})$ to the final state~$s'$ with
the energy~$\ep_{s'}$ and the wave
function~$\psi_{s'}(\vec{r})$, accompanied by absorption or
emission of one phonon, is given by the Fermi Golden Rule:
\begin{eqnarray}\nonumber
&&w^{\rm{abs(em)}}_{s\rightarrow{s}'}=
2\pi\sum_{\vec{q},\lambda}
\left|\int\psi^*_{s'}(\vec{r})\,e^{\pm{i}\vec{q}\vec{r}}
\psi_s(\vec{r})\,d^d\vec{r}\right|^2\times\\
&&\times|V_{\vec{q},\lambda}|^2
\left(N_{\vec{q},\lambda}+\frac{1}{2}\mp\frac{1}{2}\right)
\delta(\ep_{s'}-\ep_s\mp\omega_{\vec{q},\lambda}),
\end{eqnarray}
where $N_{\vec{q},\lambda}$ is the phonon occupation number
before the transition, and $\omega_{\vec{q},\lambda}$~is
the phonon frequency; the upper sign corresponds to the phonon
absorption, the lower one -- to emission.
Introducing the transition rate
\begin{equation}\label{avrate=}
w(\ep,\ep')=\delta^2\sum_{s,s'}
(w^{\rm{abs}}_{s\rightarrow{s}'}+w^{\rm{em}}_{s\rightarrow{s}'})\,
\delta(\ep-\ep_{s})\,\delta(\ep'-\ep_{s'}),
\end{equation}
averaged over the random dot realizations, we can write the
kinetic equation for the electronic distribution
function~$f_{\ep}$:
\begin{equation}\label{kinetic=}
\frac{\partial{f}_{\ep}}{\partial{t}}=
\int\left[w(\ep',\ep)\,(1-f_{\ep})f_{\ep'}
-w(\ep,\ep')\,f_{\ep}(1-f_{\ep'})\right]\frac{d\ep'}{\delta}.
\end{equation}

The average rate~(\ref{avrate=}) is determined by the electronic
wave function correlations in the dot:
\begin{eqnarray}\nonumber
\Pi_{\ep,\ep'}(\vec{r},\vec{r}')&\equiv&
\sum_{s,s'}\psi_s(\vec{r})\,\psi^*_s(\vec{r}')\,
\psi_{s'}(\vec{r}')\,\psi^*_{s'}(\vec{r})\times\\
&&\times\delta(\ep-\ep_s)\,\delta(\ep'-\ep_{s'}),
\end{eqnarray}
averaged over the dot realizations. Then we can write the average
transition rate as
\begin{eqnarray}\nonumber
w(\ep,\ep')&=&2\pi\delta^2\sum_{\vec{q},\lambda}
\Pi_{\ep,\ep'}(\vec{q},\vec{q})|V_{\vec{q},\lambda}|^2\times\\
\nonumber &&\times
\left[N_{\vec{q},\lambda}\,\delta(\ep'-\ep-\omega_{\vec{q},\lambda})
\right.+\\ &&+\left.
(N_{\vec{q},\lambda}+1)\,\delta(\ep'-\ep+\omega_{\vec{q},\lambda})
\right],
\end{eqnarray}
with the Fourier transform defined as
\begin{equation}
\Pi_{\ep,\ep'}(\vec{q},\vec{q}')\equiv
\int\Pi_{\ep,\ep'}(\vec{r},\vec{r}')\,
e^{-i\vec{q}\vec{r}+i\vec{q}'\vec{r}'}\,d^3\vec{r}\,d^3\vec{r}'.
\end{equation}
Statistical properties of ballistic dots have been extensively
studied (in Refs.~\cite{AleinerLarkin,Mirlin}, for a review see
Refs.~\cite{Mirlinrev,Aleinerrev}). For $|\ep-\ep'|$ smaller than
the Thouless energy~$E_{Th}$ one can use the following estimate:
\begin{equation}
\Pi_{\ep,\ep'}(\vec{q},\vec{q})\sim
\frac{1}{E_{Th}\delta}\,\min\{1,(q_{\|}L)^2\},
\end{equation}
where the factor $(q_{\|}L)^2$ appears when $q_{\|}L\ll{1}$
($\vec{q}_{\|}$~is the component of the wave vector parallel to
the plane of the 2DEG). As a result, the transition rate
$w(\ep,\ep')$~depends only on the transferred energy
$\omega\equiv\ep-\ep'$.

We assume the electronic temperature (determined by the
balance between heating and cooling) to be much higher than
the lattice temperature (determined by the external cryostat).
In this case one can neglect any phonon population present,
$N_{\vec{q},\lambda}=0$, so only emission of phonons can occur,
and $w(\omega)\propto\theta(\omega)$. For a power-law dependence,
$w(\omega)\propto\omega^{\alpha}\theta(\omega)$,
and Fermi-Dirac electron distribution in the
dot~(\ref{FermiDirac=}) the cooling rate is given by
\begin{equation}\label{WT=}
W_{\mathrm{out}}=
\int\frac{d\ep}{\delta}\,\frac{d\omega}{\delta}\,
\omega\,w(\omega)\,f_{\ep}(1-f_{\ep-\omega})\propto
\frac{T^{\alpha+3}}{\delta^2}\:.
\end{equation}
Obviously, such a power-law dependence can be parametrized by a
single parameter~$T_{\rm{ph}}$ and written as
$W_{\mathrm{out}}=T^{\alpha+3}/T_{\rm{ph}}^{\alpha+1}$. From the
kinetic equation~(\ref{kinetic=}) one can also extract the
single-particle relaxation rate~$\gamma_{\rm{ph}}$. For electrons
with the typical energy $\ep\sim{T}$ and
$w(\omega)\propto\omega^{\alpha}\theta(\omega)$ we obtain
$\gamma_{\rm{ph}}(T)\sim\delta\,(T/T_{\rm{ph}})^{\alpha+1}$.

\subsection{Specific mechanisms}

To consider specific electron-phonon coupling mechanisms,
we describe phonons in terms of the lattice displacement
operator for each normal phonon mode~$\lambda$:
\begin{equation}
\hat{\vec{u}}_{\lambda}(\vec{r})=\sum_{\vec{q}}
\sqrt{\frac{1}{2\mathcal{V}\rho_{\lambda}
\omega_{\vec{q},\lambda}}}\:\vec{e}_{\vec{q},\lambda}
\left[\hat{b}_{\vec{q},\lambda}e^{i\vec{q}\vec{r}}+
\hat{b}_{\vec{q},\lambda}^{\dagger}e^{-i\vec{q}\vec{r}}\right].
\end{equation}
The displacement of each mode is directed along the unit vector
$\vec{e}_{\vec{q},\lambda}$.
To each mode corresponds some mass which is the total mass of the
unit cell for acoustic phonons or the reduced mass for optical
phonons; dividing it by the unit cell volume one obtains the
corresponding density~$\rho_{\lambda}$. Finally, $\mathcal{V}$~is
the 3D quantization volume.
At low temperatures we are interested in, only acoustic phonons
can be emitted. We approximate their dispersion by
$\omega_{q}=v_sq$, with $v_s$~being the sound velocity, while
the density~$\rho_{\lambda}$ coincides with the density of the
crystal~$\rho_0$.

{\em Deformational coupling} to the acoustic phonons is due to
the local change of the electronic energy bands under strain:
\begin{equation}
\hat{V}^{\rm{def}}(\vec{q})=
\Xi_{jl}\,iq_j\hat{u}_l(\vec{q}),
\end{equation}
where $\Xi_{jl}$~is the deformational coupling tensor. In a bulk
crystal it, generally speaking, depends on the electronic wave
vector~$\vec{k}$~\cite{Levinson}. In doped GaAs, when typical
electronic wave vectors are close to the Brillouin zone center
and one approximates the periodic part of the Bloch function by
that for $\vec{k}=0$, this dependence vanishes and
$\Xi_{jl}=\Xi\delta_{jl}$. The leading anisotropic (i.~e.,
dependent on the direction of~$\vec{k}$) correction at small
finite~$\vec{k}$ should be smaller by a factor of $(ka)^2$,
where $a$~is the lattice constant (the first-order in~$\vec{k}$
correction should vanish due to the time-reversal symmetry).
Hence we can estimate its magnitude as
$\sim(k_Fa)^2\Xi\sim{n}a^2\Xi$, where $k_F$~is the electronic
Fermi wave vector, and $n$~is the 2D~electron density.

The isotropic (independent of the direction of $\vec{k}$) part of
the deformation potential is subject to screening~\cite{Levinson}.
The electrons inside the dot can screen the fields with wave
vectors $q_{\|}$ down to~$\sim{1}/L$. We assume that the Fourier
components with $q_{\|}\ll{1}/L$ are also screened, either by the
2DEG outside the dot, or by the metallic gate.
Thus, by the order of magnitude, we can use the expression
for the static (due to $v_s\ll{v}_F$) screening by an infinite
2DEG, which results in the renormalization:
\begin{equation}
\Xi\rightarrow\frac{\Xi}{1+1/(q_{\|}a_s)}\approx
q_{\|}a_s\Xi\:,
\end{equation}
where $a_s$ is the 2D screening length (equal to half the
electronic Bohr radius), and we consider $q_{\|}a_s\ll{1}$.
Thus, the effective deformation potential is suppressed by a
small factor: either by $q_{\|}a_s$ for the isotropic part of
the potential, or by $na^2$ for the anisotropic part.

{\em Piezoelectric coupling} to acoustic phonons is due to the
longitudinal electric field induced by the strain. We express
the potential in terms of the electromechanical
tensor~$e^{\rm{em}}_{ijl}$, which relates the induced
polarization to the strain tensor:
\begin{equation}
\hat{V}^{\rm{piezo}}(\vec{q})=
-\frac{4\pi{e}{e}^{\rm{em}}_{ijl}}{\vep}\,
\frac{q_iq_j}{q^2}\,\hat{u}_l(\vec{q}).
\end{equation}
Here $\vep$~is the background dielectric constant of the
material. The in-plane piezoelectric field is also subject to
screening, which brings a factor of $q_{\|}a_s$.

The component of the piezoelectric field perpendicular to
the dot plane is not screened by the electrons. Instead, it
affects the confinement and shifts the subbands, which can
be viewed as Stark effect. If $q_{\|}\neq{0}$, the shift of
the subband depends on the in-plane coordinate and represents
an additional effective potential felt by the electrons.
If we assume the confinement of the electrons by an asymmetric
triangular potential well formed by the constant force~$F$ on
one side and a hard wall on the other, the confinement energy
$\ep_z\sim(\hbar^2F^2/m)^{1/3}$, while
$\partial\ep_z/\partial{F}\equiv{a}_z$ is of the order of the
extent of the confined state in the $z$~direction.
Thus, we can estimate
\begin{equation}
\hat{V}^{\rm{Stark}}(\vec{q})\sim
q_za_z\hat{V}^{\rm{piezo}}(\vec{q})\:.
\end{equation}
This effective in-plane potential is also subject to screening,
which brings an additional factor of $q_{\|}a_s$.

As a result, we can generally write
\begin{equation}
V_{\vec{q}}\sim{A}\sqrt{\frac{q}{\mathcal{V}\rho_0v_s}}\:,
\end{equation}
with~$A$ given by
\[
qa_s\Xi,\;\;na^2\Xi,\;\;\frac{4\pi{e}e^{\rm{em}}a_s}{\vep}\:,\;\;
q_za_z\,\frac{4\pi{e}e^{\rm{em}}a_s}{\vep}\:,
\]
for the screened isotropic deformation potential, anisotropic
deformation potential, screened in-plane piezoelectric field,
and the perpendicular piezoelectric field, respectively.

Let us estimate the relative importance of these mechanisms,
using the numbers for GaAs from Ref.~\cite{LB}. The bare
deformation potential $\Xi\sim{10}$~eV, the screening length
$a_s\approx{50}$~{\AA}, the lattice constant
$a\approx{5}$~{\AA}.
We will be interested in temperatures $T\sim{0.1}$--$1$~K,
so we indeed have $qa_s\ll{1}$, and we are in the regime
$qL\gg{1}$ (for $L\sim{1}$~$\mu$m). For $v_sq=1$~K we have
$q\approx{3}\cdot{10}^{-3}$~{\AA}$^{-1}$
($v_s\approx{5}\cdot{10}^5$~cm/s), so $qa_s\approx{0}.15$.
For $n=10^{12}$~cm$^{-2}$ $na^2\approx{2.5}\cdot{10}^{-3}$,
so the screened isotropic part is more important than the
anisotropic one.
The only independent component of the electromechanical tensor
in GaAs $e^{\rm{em}}_{14}\approx{1.4}\cdot{10}^7$~V/cm,
the dielectric constant $\vep\approx{13}$, so
$4\pi{e}e^{\rm{em}}_{14}a_s/\vep\approx{7}$~eV, which is of
the order of the unscreened deformation potential. For the
screened potential due to the perpendicular piezoelectric
field, as typically $a_z\sim{100}$~{\AA}~\cite{Davies}, we
have a smallness of $q_z{a}_z$.
In conclusion, contrary to the estimates of
Ref.~\cite{Galperin}, we obtain that the in-plane piezoelectric
coupling is more important than the deformational one.

As a result, we arrive at the estimate
\begin{equation}\label{Woutph=}
W_{\rm{out}}(T)\sim\frac{(4\pi{e}e^{\rm{em}}a_s/\vep)^2}{\rho_0v_s^5E_{Th}
\delta}\,T^6\equiv\frac{T^6}{T_{\rm{ph}}^4}\:,
\end{equation}
For GaAs $\rho_0v_s^5\approx(0.074$~eV$)^4$,
(the density $\rho_0\approx{5.3}$~g/cm$^3$),
for a typical dot~\cite{Marcus} $\delta\sim{1}$~$\mu$eV,
$E_{Th}\sim{100}$~$\mu$eV, so we obtain
$T_{\rm{ph}}\sim{0.1}$~meV~$\sim{1}$~K.

\section{Stationary state}

\subsection{Ohmic absorption}

First, consider the case of the simple Ohmic absorption with
cooling only due to the contacts in the sequential tunneling
regime with the rate given by Eq.~(\ref{WoutCBP=}).
At small detunings ($U\ll{T}$) we have
\begin{equation}
W_{\rm{out}}(T)=\frac{\gamma}{\delta}
\left[\frac{\pi^2T^2}{12}-\frac{U^2}{4}+O(U^4)\right],
\end{equation}
so that the stationary temperature is given by
\begin{equation}
T(U)=\frac{2}{\pi}\sqrt{\frac{3W_0}{\gamma/\delta}}
\left[1+\frac{(\gamma/\delta){U}^2}{8W_0}+O(U^4)\right].
\end{equation}
The temperature $T(U=0)$ determines the curvature of the Coulomb
blockade peak at $U=0$: from Eq.~(\ref{GVG0=}) we have
\begin{equation}\label{GU0=}
\frac{G(U)}{G_0}=1-\frac{1}{4\ln{2}}\,\frac{U^2}{T^2(U=0)}\:.
\end{equation}

At large detunings $\gg{T}$ we can approximate the right-hand
side of Eq.~(\ref{WoutCBP=}) by $|x|e^{-2|x|}$ and write
\begin{equation}\label{TwingCl=}
\frac{W_0}{(\gamma/\delta)T^2}\approx|x|e^{-2|x|}\:,\quad
T\approx\frac{U}{\ln[(\gamma/\delta)U^2/(2W_0)]}\:,
\end{equation}
with the logarithmic precision. It is correct if the logarithm
in the denominator is large, or $U\gg{T}(U=0)$.
This result means that the tails of the Coulomb blockade peak
have the form:
\begin{equation}\label{gV2=}
\frac{G(U)}{G_0}\approx\frac{2W_0}{(\gamma/\delta)U^2}
\ln\left[\frac{(\gamma/\delta)U^2}{2W_0}\right].
\end{equation}
The weak power-law fall-off of the tails is drastically different
from the exponential one occurring in equilibrium:
$G(U)/G_0=(U/T)/\sinh(U/T)$~\cite{Shekhter}. The reason
for this difference is very simple: as the gate voltage is tuned
away from the degeneracy point, the exchange of electrons between
the dot and the contacts becomes weaker, so the cooling rate
decreases leading to an increase in the temperature and hence in
the conductance.

At large enough detunings the cooling becomes dominated by the
inelastic cotunneling, Eq.~(\ref{WoutCBV=}), rather then sequential
tunneling, Eq.~(\ref{WoutCBP=}). In this regime the dot temperature
and the conductance are given by
\begin{equation}\label{gV=}
T=\left(\frac{15}{2\pi^3}\,\frac{W_0U^2}{(\gamma/\delta)^2}\right)^{1/4},
\quad
\frac{G(U)}{G_0}=\sqrt{\frac{10}{3\pi^2}}\,\frac{\sqrt{W_0}}{U}\:.
\end{equation}
The switching to the inelastic cotunneling occurs at
\begin{equation}
U\sim\sqrt{W_0}\,\frac{\delta}{\gamma}\ln^2\frac{\delta}{\gamma}\:,\quad
\frac{G(U)}{G_0}\sim\frac{\gamma/\delta}{\ln^2(\delta/\gamma)}\:.
\end{equation}
The logarithmic precision of these estimates, however, makes them
applicable only for extremely small~$\gamma$ [such that
$\ln(\delta/\gamma)\gg{1}$]. In reality, if one takes directly the
expressions~(\ref{WoutCBP=}) and~(\ref{WoutCBV=}) for the cooling
rate, for $W_0=30$~$\mu$eV$^2\approx{46}$~$\mu$eV/s,
$\gamma/\delta=0.2$~\cite{numbers} the contribution of the inelastic
cotunneling starts to affect the stationary electronic temperature
noticeably [as compared to the precision of Eq.~(\ref{TwingCl=})]
only as far as $V>1$~meV. At $V=1$~meV the conductance
$G(1$~meV$)/G_0\approx{0.01}$, and about 18\%~of it is still due
to the sequential tunneling.

\begin{figure}
\psfig{figure=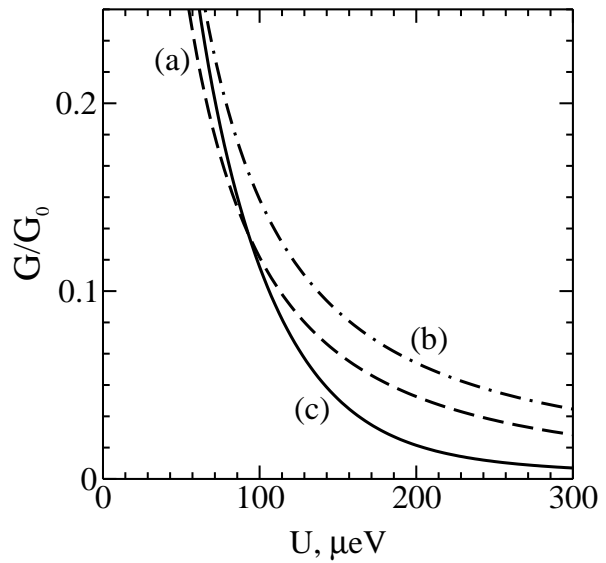,width=9cm,height=9cm}
\caption{\label{CBPclcot:}
Normalized conductance versus reduced gate voltage
(Coulomb blockade tail):
(a)~only sequential tunneling is taken into account,
(b)~cotunneling is added, and
(c)~cooling by phonons is present as well~\cite{numbers}.}
\end{figure}

\begin{figure}
\psfig{figure=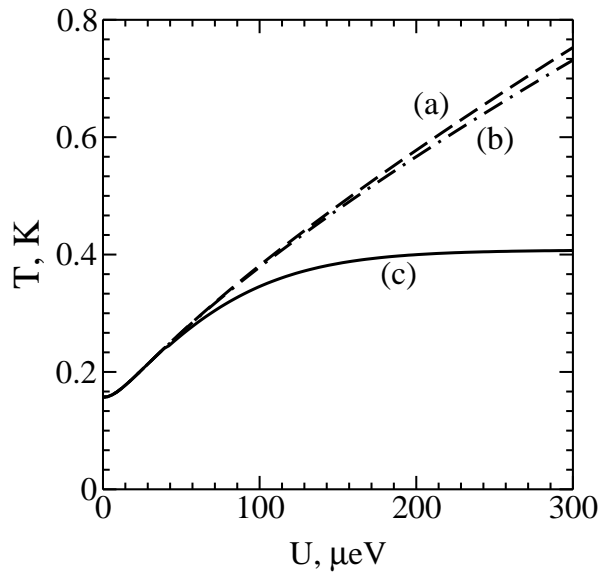,width=9cm,height=9cm}
\caption{\label{Tcl:}
Electronic temperature in kelvins versus reduced gate voltage:
(a)~only sequential tunneling is taken into account,
(b)~cotunneling is added, and
(c)~cooling by phonons is present as well~\cite{numbers}.}
\end{figure}

If one takes now into account cooling by phonons with the
rate~(\ref{Woutph=}), it sets the upper limit for the electronic
temperature: $T_{\rm{max}}=(W_0T_{\rm{ph}}^4)^{1/6}$. If the pumping
is strong enough (or the dot is closed enough),
$T_{\rm{max}}\ll\sqrt{W_0\delta/\gamma}$, the phonon cooling
mechanism dominates, the dot temperature is constant and equal
to~$T_{\rm{max}}$ for all~$U$, so that the Coulomb blockade peak shape
is given explicitly by Eq.~(\ref{GVG0=}), and its tails -- by
Eq.~(\ref{WoutCBV=}).
In the opposite limiting case,
$T_{\rm{max}}\gg\sqrt{W_0\delta/\gamma}$ the electronic temperature
in the peak region is determined by electron escape, and only in the
peak tails, when the dot effectively becomes more and more closed,
phonon emission starts to dominate. This will manifest itself as a
crossover from the $1/U$ tail~(\ref{gV=}) to the $1/U^2$ one
given by Eq.~(\ref{WoutCBV=}) at fixed $T=T_{\rm{max}}$. This
crossover occurs at $U\sim(\gamma/\delta){T}_{\rm{max}}^2/\sqrt{W_0}$,
$G/G_0\sim(\delta/\gamma)W_0/T_{\rm{max}}^2$~\cite{crossover}.
We plot the tails of the Coulomb blockade peak $G(U)/G_0$ in
Fig.~\ref{CBPclcot:} for three cases when (a)~only sequential
tunneling is taken into account, (b)~cotunneling is added, and
(c)~cooling by phonons is present as well~\cite{numbers}.
In Fig.~\ref{Tcl:} we plot the electronic temperature $T(U)$ for
the same three cases.

\subsection{Dynamic localization}\label{DynLoc}

As we have discussed in Sec.~\ref{DL}, in the strong dynamic
localization regime the residual absorption is determined by
dephasing. Using the results of the previous sections we can
identify three sources of dephasing.

(i) {\it Escape to the contacts}.
The quasiparticle relaxation rates for the sequential
tunneling and inelastic cotunneling are given by
Eqs.~(\ref{gescCBP=}) and~(\ref{WoutCBV=}).

(ii) {\it Phonon emission.} According to the arguments given
in the end of Sec.~\ref{PhononsGeneral}, we can write
\begin{equation}
\gamma_{\rm{ph}}(T)\sim\delta\left(\frac{T}{T_{\rm{ph}}}\right)^4.
\end{equation}

(iii) {\it Electron-electron collisions.} The corresponding
quasiparticle relaxation rate in a quantum dot was calculated
by Sivan, Imry and Aronov~\cite{SIA}:
\begin{equation}
\gamma_{e-e}(T)\sim\delta\left(\frac{T}{E_{Th}}\right)^2,
\end{equation}
where $E_{Th}$~is the Thouless energy. The derivation of this
expression implies the effective continuity of the many-particle
spectrum, which imposes a condition
$T_*\gg\sqrt{E_{Th}\delta/\ln(E_{Th}/\delta)}$~\cite{AGKL}.
Obviously, for the dynamic localization to have any chance to
develop, the condition $\gamma_{\phi}(T_*)t_*\ll{1}$ should be
satisfied.

\begin{figure}
\psfig{figure=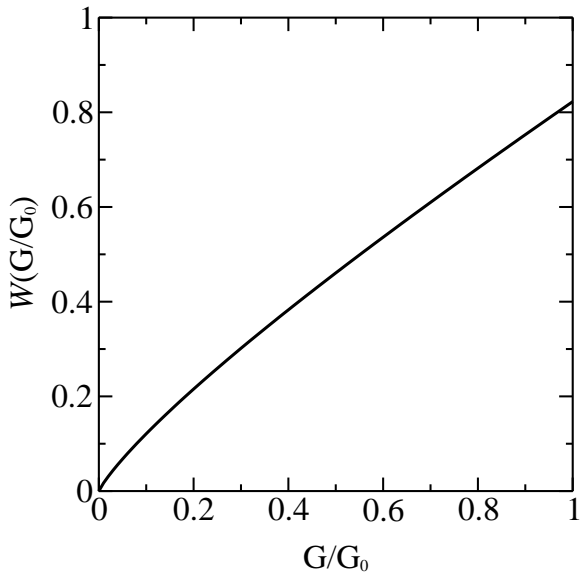,width=9cm,height=9cm}
\caption{\label{calW:}
The function $\mathcal{W}(G/G_0)$ defined in Eq.~(\ref{calW=}),
for sequential tunneling.}
\end{figure}

Suppose for a moment that dephasing is dominated by
electron-electron collisions, while cooling is dominated
by the escape to the contacts (later we will analyze the
conditions for this to be true).
One can notice a common property of Eqs.~(\ref{GVG0=}),
(\ref{WoutCBP=}), and (\ref{WoutCBV=}): for both sequential
tunneling and cotunneling $G/G_0$ and
$W_{\rm{out}}(U)/[(\gamma/\delta)T^2]$ are functions of
$x\equiv{U}/(2T)$ only. This allows us to write a relation
\begin{equation}\label{calW=}
W_{\rm{out}}=(\gamma/\delta)T^2\,\mathcal{W}(G/G_0)\,.
\end{equation}
The energy balance condition takes the form
\begin{equation}
W_{\rm{in}}\sim{T}_*^2\,\frac{T^2}{E_{Th}^2}=
W_{\rm{out}}=\frac{\gamma}{\delta}\,T^2\,\mathcal{W}(G/G_0)\,,
\end{equation}
or $(\gamma/\delta)\mathcal{W}(G/G_0)=(T_*/E_{Th})^2$. Since
$U$~and~$T$ have dropped out, the solution of this equation
for~$G$ is independent of~$U$, leading to a flat plateau
on the Coulomb blockade curve $G(U)$~\cite{BK}. With logarithmic
precision $\mathcal{W}(G/G_0)\sim{G}/G_0$ (see Fig.~\ref{calW:}),
so the level of the plateau is
$G/G_0\sim(\delta/\gamma)(T_*/E_{Th})^2$.

Note that the largest possible value of $\mathcal{W}(G/G_0)$
is $\pi^2/12$ reached at $G/G_0=1$ (corresponding to $U=0$).
Therefore, the solution exists only if
\begin{equation}\label{conditionopen=}
\frac{\gamma}{\delta}\gg
\left(\frac{T_*}{E_{Th}}\right)^2.
\end{equation}
Physically, this means that the dot should be sufficiently
open, so that the cooling is intense enough and the stationary
temperature is not too high to destroy the localization.
Note that for the observation of the plateau the condition
$\gamma\ll{1}/t_*$ {\em is not} necessary: even if at $U=0$
the dynamic localization is absent, as $U$~is increased, the
dot becomes effectively more closed, so the dephasing by escape
becomes less efficient.
Of course, for the Coulomb blockade itself to be present,
the condition $\gamma/\delta\ll{1}$ should be
satisfied~\cite{contacts}.

Now let us consider the very top of the peak, $U=0$. Including
the dephasing due to both escape and electron-electron collisions,
we can write the energy balance condition as
\begin{equation}\label{balanceU0=}
\frac{\gamma}{\delta}\,T^2\sim\frac{\gamma}{\delta}\,T_*^2+
\frac{T^2}{E_{Th}^2}\,T_*^2\:.
\end{equation}
Here the left-hand side represents the cooling rate in the peak,
the first term on the right-hand side comes from the dephasing
due to escape, and the second term represents the contribution
from collisions. Due to the condition~(\ref{conditionopen=})
the second term is necessarily negligible compared to the
left-hand side, so the only way to satisfy the equation is to
have $T(U=0)\sim{T}_*$. Thus, for the dynamic localization to be
possible the dephasing in the very peak of the Coulomb blockade
{\em must} be dominated by escape.

\begin{figure}
\psfig{figure=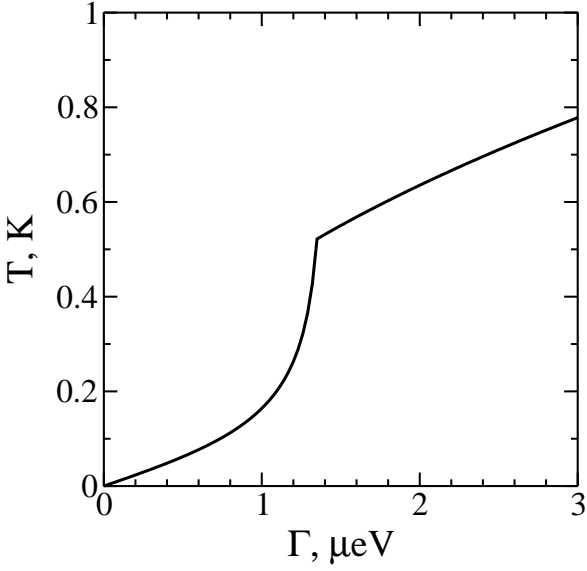,width=9cm,height=9cm}
\caption{\label{TU0:}
Dependence of the electronic temperature in the peak $T(U=0)$ on
the perturbation intensity~$\Gamma$ for $\gamma/\delta=0.02$,
$\delta=0.3$~$\mu$eV, $\omega=3$~$\mu$eV, $E_{Th}=100$~$\mu$eV;
in reality the sharp angle is replaced by a smooth crossover.}
\end{figure}

Eq.~(\ref{GU0=}) remains valid in the dynamic localization regime
as well, as it does not depend on the details of heating and
cooling mechanisms. Thus, one can extract the temperature of the
stationary state at $U=0$ measuring the curvature of the peak,
and study its dependence on control parameters: intensity~$\Gamma$
and coupling to the contacts~$\gamma$. From Eq.~(\ref{balanceU0=})
it is seen that this dependence is the strongest when
$\gamma/\delta$ is close to $T_*^2/E_{Th}^2$ (up to a numerical
coefficient), i.~e. when the dynamic localization in the peak is
about to be destroyed.
If we plot $T(U=0)$ versus~$\Gamma$ (Fig.~\ref{TU0:}), we see
that destruction of the dynamic localization manifests itself
as a crossover from the linear dependence $T\propto\Gamma$ deep
in the DL regime (small~$\Gamma$) to $T\propto\sqrt{\Gamma}$
in the Ohmic regime. According to the abovesaid, this crossover
can be quite pronounced (like shown in the figure) when
\begin{equation}\label{DLdestroy=}
\gamma\sim\delta\,\frac{T_*^2}{E_{Th}^2}\ll\frac{1}{t_*}\:.
\end{equation}

\begin{figure}
\psfig{figure=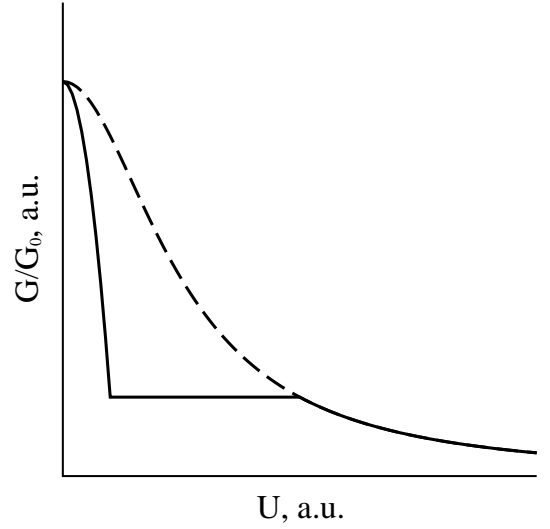,width=9cm,height=9cm}
\caption{\label{CBscheme:}
A sketch of the Coulomb blockade peak shape in the dynamic
localization regime without taking into account the phonon
cooling (solid line): at small~$U<U_{\rm{min}}$ the dephasing
is dominated by the electron escape (peak), at larger~$U$ --
by electron-electron collisions (plateau), and finally, at
$U>U_{\rm{max}}$ the cooling is insufficient, the dynamic
localization is destroyed, and the dot is in the Ohmic regime.
The Ohmic curve is also shown for reference by the dashed line.}
\end{figure}

As $U$~is detuned away from the peak, the dot becomes effectively
more closed, and simultaneously the electronic temperature grows
and electron-electron collisions become more frequent. Thus, the
crossover from the peak to the plateau occurs where the two
mechanisms are equally efficient.
With the logarithmic precision this happens at
\begin{equation}\label{beginplateau=}
T\sim{T}_*\,\quad U\sim{U}_{\rm{min}}\sim{T}_*
\max\left\{1,\frac{\gamma}{\delta}\,\frac{E_{Th}}{T_*}\right\},
\end{equation}
depending on whether the plateau is in the region of sequential
tunneling, $\gamma/\delta\ll{T_*}/E_{Th}$, or of inelastic
cotunneling, $\gamma/\delta\gg{T_*}/E_{Th}$. The plateau ends
when the temperature of the dot becomes so large that the dynamic
localization is destroyed by dephasing. Obviously, this happens
when the horisontal line $G/G_0=(\delta/\gamma)(T_*/E_{Th})^2$
hits the curve~(\ref{gV2=}) or~(\ref{gV=}), which happens at
\begin{equation}\label{endplateau=}
U\sim{U}_{\rm{max}}\sim{E}_{Th}\sqrt{\frac{\delta}{\Gamma}}
\max\left\{1,\frac{\gamma}{\delta}\,\frac{E_{Th}}{T_*}\right\}.
\end{equation}
The resulting shape of the Coulomb blockade peak is drawn
schematically in Fig.~\ref{CBscheme:} for the Ohmic absorption
and dynamic localization regimes.
The two boundaries (\ref{beginplateau=})~and~(\ref{endplateau=})
give a nonzero range of~$U$ (i.~e. $U_{\rm{min}}<U_{\rm{max}}$),
if $T_*\ll{E}_{Th}\sqrt{\delta/\Gamma}$, which can be equivalently
rewritten as $\gamma_{e-e}(T_*)\ll{1}/t_*$, i.~e. a necessary
condition for the dynamic localization itself.

\begin{figure}
\psfig{figure=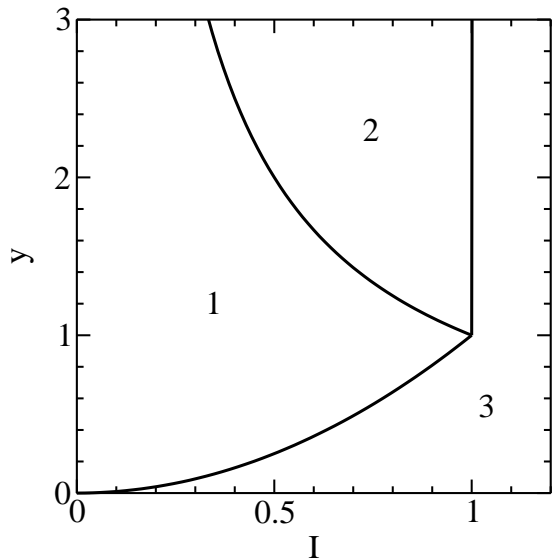,width=9cm,height=9cm}
\caption{\label{phasediag:}
A schematic view of the ``phase diagram'' in terms of the
dimensionless intensity and escape rate ($I-y$ plane), without
taking into account cooling and dephasing due to phonons.
The top of the Coulomb blockade peak corresponds to dynamic
localization regime only in the region~1; the flat plateau
in the tails exists both in  regions $1$~and~$2$; in the
region~3 DL is absent.}
\end{figure}

It is convenient to introduce two dimensionless parameters,
corresponding to two experimentally controllable parameters
$\Gamma$~and~$\gamma$:
\begin{equation}
I\equiv\frac{\Gamma}{\delta}
\left(\frac{\omega}{E_{Th}}\right)^{2/3},\quad
y\equiv\frac{\gamma}{\delta}
\left(\frac{\omega}{E_{Th}}\right)^{-2/3}.
\end{equation}
The condition $\gamma_{e-e}(T_*)t_*\ll{1}$ becomes $I\ll{1}$,
the condition~(\ref{conditionopen=}) is $y\gg{I}^2$. The top
of the peak will correspond to DL regime if $\gamma{t}_*\ll{1}$
or $Iy\ll{1}$. The resulting ``phase diagram'' is shown in
Fig.~\ref{phasediag:}. The conditions~(\ref{DLdestroy=})
for a pronounced crossover in Fig.~\ref{TU0:} correspond to
crossing the parabola $y=I^2$ in its lower part.

So far, when analyzing the dynamic localization, we did not take
phonons into account.
Now consider another extreme case: both cooling and dephasing
are entirely due to phonons. Then the energy balance condition
in the localization regime reads as
\begin{equation}
W_{\rm{in}}\sim{T}_*^2\,\frac{T^4}{T_{\rm{ph}}^4}
=\frac{T^6}{T_{\rm{ph}}^4}=W_{\rm{out}}\,,
\end{equation}
giving $T\sim{T}_*$. Note that this conclusion is independent of the
power of temperature in the phonon cooling rate [or of~$\alpha$
appearing in Eq.~(\ref{WT=})]. Obviously, phonons will dominate if
$T_{\rm{ph}}\ll{T_*}(\delta/\gamma)^{1/4}$,
$T_{\rm{ph}}\ll\sqrt{T_*E_{Th}}$. In this case the shape of the
peak is given explicitly by Eq.~(\ref{GVG0=}), its tails -- by
Eq.~(\ref{WoutCBV=}), and the width corresponds to the electronic
temperature of the dot. The signature of the dynamic
localization effect would be the linear dependence of the temperature
on the microwave power, in contrast to the $1/6$~power for the
Ohmic absorption case (see the previous subsection). The localization
regime exists as long as $\gamma_{\rm{ph}}(T_*)\ll{1}/t_*$, or
$T_*^5\ll\omega{T}_{\rm{ph}}^4$. The solution for the Ohmic regime is
$T_{\rm{max}}=(\omega{T}_*T_{\rm{ph}}^4)^{1/6}$, and it is stable as
long as $\gamma_{\rm{ph}}(T_{\rm{max}})\gg{1}/t_*$, which gives
$T_*^5\gg\omega{T}_{\rm{ph}}^4$. Thus, at a certain intensity such
that $T_*\sim(\omega{T}_{\rm{ph}}^4)^{1/5}$ there is a crossover
between the localization and Ohmic regimes.

\begin{figure}
\psfig{figure=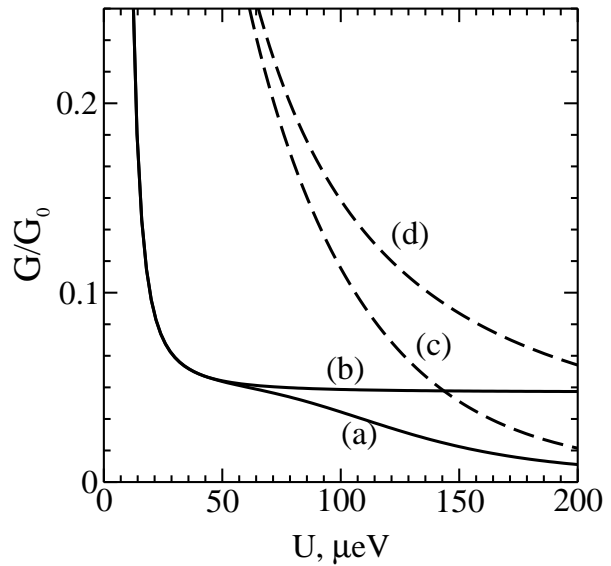,width=9cm,height=9cm}
\caption{\label{CBPloc:}
Normalized conductance versus reduced gate voltage
(Coulomb blockade tail):
dynamic localization regime with (curve~a) and without (curve~b)
phonon cooling taken into account, and the same for the purely Ohmic
absorption (curves c~and~d)~\cite{numbers}.}
\end{figure}

Including all mechanisms, we can note that if the electron-phonon
interaction is weak enough, $T_{\rm{ph}}^2\gg{E}_{Th}^3\omega/T_*^2$,
the phonon cooling plays any role only in the Ohmic part of the
Coulomb blockade tail. Otherwise, phonons start to ``eat up'' the
plateau from the large~$U$ side~\cite{cooldeph}. The plateau will
disappear at $T_{\rm{ph}}\sim\sqrt{T_*E_{Th}}$. As an illustration,
for the intermediate case, we plot the Coulomb blockade tail in
Fig.~\ref{CBPloc:} in the dynamic localization regime with and
without phonon cooling (lower and upper solid curves, respectively)
together with the corresponding Ohmic curves shown by dashed lines.

\section{Conclusions}

We have studied electronic conduction through a quantum dot in
the Coulomb blockade regime under an external periodic perturbation.
In contrast to the well-studied equilibrium case, the electronic
temperature of the dot under pumping is different from that of
the contacts and the substrate. It is determined by the balance
between heating by the perturbation and cooling due to electron
exchange with contacts and phonon emission. When the cooling is
dominated by the former mechanism, its rate depends on the gate
voltage, and so does the dot temperature. As the gate voltage is
detuned away from the peak, the cooling rate decreases, and the
temperature increases. As a result, the tails of the Coulomb
blockade peak fall off less rapidly than in the equilibrium case:
instead of the usual exponential fall-off for the sequential
tunneling, under pumping one has a power-law
dependence~(\ref{gV2=}), while for the inelastic cotunneling
the equilibrium power law is replaced by a weaker one,
Eq.~(\ref{gV=}).
At sufficiently high temperatures cooling by phonons becomes
important, which sets an upper limit for the dot temperature
(depending on the pumping intensity), which however, can be
significantly higher than the cryostat temperature.

In the strong dynamic localization regime the heating rate is
determined by dephasing, as the usual linear absorption is
blocked by quantum interference. The dephasing can be due to
electron-electron collisions, electron escape to the contacts,
as well as phonon emission. The most peculiar situation is
realized when the cooling is due to the contacts, while the
dephasing is due to electron-electron collisions: in this case
the Coulomb blockade peak has a flat shoulder, where the
conductance does not depend on the gate voltage. Such a shape
could be an experimental signature of the dynamic localization
effect.

Finally, we wish to note that conductance measurements are not
necessarily the only possible way to detect the dynamic
localization.
An isolated mesoscopic sample can be put into a microwave cavity,
and the energy absorption rate can be measured as it affects the
$Q$-factor of the cavity~\cite{Bouchiat}. In this case the only
cooling mechanism is phonon emission, while the dephasing can be
due to electron-electron interactions as well. In the dynamic
localization regime the absorption rate depends nonlinearly on the
ac field intensity: $W_{\rm{in}}\propto\Gamma^3$ if the
dephasing is dominated by electron-electron collisions
($T_{\rm{ph}}\gg\sqrt{T_*E_{Th}}$), or
$W_{\rm{in}}\propto\Gamma^6$ if the dephasing is dominated by
phonons ($T_{\rm{ph}}\ll\sqrt{T_*E_{Th}}$).

\section*{Acknowledgments}

The authors are grateful to Yu.~M.~Galperin, C.~M.~Marcus,
B.~L.~Altshuler, V.~I.~Fal'ko, and B.~N.~Narozhny for helpful
discussions.


\begin{thebibliography}{99}


\bibitem{MarcusRev}
L.~P.~Kouwenhoven, C.~M.~Marcus, P.~L.~McEuen, S.~Tarucha,
R.~M.~Westervelt, and N.~S.~Wingreen,
in {\it Mesoscopic Electron Transport}, edited by
L.~L.~Sohn, L.~P.~Kouwenhoven, and G.~Sch\"on
(Kluwer, Dordrecht, 1997).

\bibitem{Aleinerrev}
I.~L.~Aleiner, P.~W.~Brouwer, and L.~I.~Glazman,
Phys. Rep. {\bf 358}, 309 (2002).

\bibitem{Marcus}
A.~G.~Huibers, J.~A.~Folk, S.~R.~Patel, C.~M.~Marcus,
C.~I.~Duru\"oz, and J.~S.~Harris,~Jr.,
Phys. Rev. Lett. {\bf 83}, 5090 (1999);
L.~DiCarlo, C.~M.~Marcus, and J.~S.~Harris,~Jr.,
Phys. Rev. Lett. {\bf 91}, 246804 (2003).

\bibitem{Raizen}
F.~L.~Moore,
J.~C.~Robinson, C.~Bharucha, P.~E.~Williams, and M.~G.~Raizen,
Phys. Rev. Lett. {\bf 73}, 2974 (1994).

\bibitem{Izrailev}
F.~M.~Izrailev, Phys. Rep. {\bf 196}, 299 (1990).

\bibitem{Haake}
F.~Haake, {\it Quantum signatures of chaos}
(Springer-Verlag, Berlin, 2001).

\bibitem{us}
D.~M.~Basko, M.~A.~Skvortsov, and V.~E.~Kravtsov,
Phys. Rev. Lett. {\bf 90}, 096801 (2003).

\bibitem{Vavilov}
M.~G.~Vavilov and I.~L. Aleiner, Phys. Rev.~B
{\bf 64}, 085115 (2001).

\bibitem{Kanzieper}
V.~I.~Yudson, E.~Kanzieper, and V.~E.~Kravtsov, Phys. Rev.~B {\bf 64},
045310 (2001).

\bibitem{perturbation}
The transition rate is proportional to the power of the microwave
and can be estimated as $\Gamma\sim(e\mathcal{E}L)^2/E_{Th}$,
where $\mathcal{E}$~is the amplitude of the electric field in
the dot, $L$~is the dot size, and $E_{Th}$ is the Thouless energy.
If the screening length~$a_s<L$, one should substitute
$\mathcal{E}L\rightarrow\mathcal{E}_{ext}a_s$.
Finally, instead of using microwave one can change the dot shape
modulating the gate voltage.

\bibitem{BK}
D.~M.~Basko and V.~E.~Kravtsov, Phys. Rev. Lett. (to be published).

\bibitem{Basko}
D.~M.~Basko, Phys. Rev. Lett. {\bf 91}, 206801 (2003).

\bibitem{Shekhter}
I.~O.~Kulik and R.~I.~Shekhter,
Zh. Eksp. Teor. Fiz. {\bf 68}, 623 (1975)
[Sov. Phys. JETP {\bf 41}, 308 (1975)].

\bibitem{Kouwenhoven2003}
R.~Deblock, E.~Onac, L.~Gurevich, and L.~P.~Kouwenhoven,
Science {\bf 301}, 203 (2003).

\bibitem{Averin}
D.~V.~Averin and Yu.~N.~Nazarov,
Phys. Rev. Lett. {\bf 65}, 2446 (1990).

\bibitem{GlazmanMatveev}
L.~I.~Glazman and K.~A.~Matveev,
Pis'ma Zh. Eksp. Teor. Fiz. {\bf 51}, 425 (1990)
[JETP Lett. {\bf 51}, 484 (1990)].

\bibitem{TienGordon}
P.~K.~Tien and J.~P.~Gordon, Phys. Rev. {\bf 129}, 647 (1963).

\bibitem{Yudson}
V.~I.~Yudson and V.~E.~Kravtsov,
Phys. Rev.~B {\bf 67}, 155310 (2003).

\bibitem{Altshuler}
F.~Zhou, B.~Spivak, N.~Taniguchi, and B.~L.~Altshuler,
Phys. Rev. Lett. {\bf 77}, 1958 (1996).

\bibitem{Galperin}
Y.~M.~Galperin and K.~A.~Chao,
Found. Phys. {\bf 30}, 2135 (2000).

\bibitem{Reizer}
M.~Yu.~Reizer and A.~V.~Sergeev,
Zh. Eksp. Teor. Phys. {\bf 90}, 1056 (1986);
{\it ibid.} {\bf 92}, 2291 (1987)
[Sov. Phys. JETP {\bf 63}, 616 (1986);
{\it ibid.} {\bf 65}, 1291 (1987)].

\bibitem{AleinerLarkin}
I.~L.~Aleiner and A.~I.~Larkin, Phys. Rev.~B {\bf 54}, 14423 (1996).

\bibitem{Mirlin}
Ya.~M.~Blanter, A.~D.~Mirlin, and B.~A.~Muzykantskii,
Phys. Rev. Lett. {\bf 80}, 4161 (1998).

\bibitem{Mirlinrev}
A.~D.~Mirlin, Phys. Rep. {\bf 326}, 259 (2000).

\bibitem{Levinson}
V.~F.~Gantmakher and Y.~B.~Levinson, {\it Carrier Scattering in
Metals and Semiconductors} (North Holland, Amsterdam, 1987).

\bibitem{LB}
{\it Landolt-B\"ornstein Numerical Data and Functional Relationships
in Science and Technology, New Series}, v.~III/17a, ed.~by O.~Madelung
(Springer, Berlin, 1982).

\bibitem{Davies}
J.~H.~Davies, {\it The Physics of Low-Dimensional Semiconductors}
(Cambridge University Press, 1997).

\bibitem{numbers}
The parameters used for the plots in Figs. \ref{CBPclcot:},
\ref{Tcl:}, and \ref{CBPloc:} are the following:
$\delta=0.3$~$\mu$eV, $\Gamma=1$~$\mu$eV, $\omega=3$~$\mu$eV
($\omega/2\pi\approx$~0.7~GHz), $E_{Th}=100$~$\mu$eV,
$T_{\rm{ph}}=1$~K~$\approx{86}$~$\mu$eV.

\bibitem{crossover}
It is worth noting that due to the condition
$T_{\rm{max}}\gg\sqrt{W_0\delta/\gamma}$ the crossover to phonon
cooling occurs after the crossover from sequential tunneling to
cotunneling.

\bibitem{SIA}
U.~Sivan, Y.~Imry, and A.~G.~Aronov,
Europhys. Lett. {\bf 28}, 115 (1994).

\bibitem{AGKL}
B.~L.~Altshuler, Y.~Gefen, A.~Kamenev, and L.~S.~Levitov,
Phys. Rev. Lett. {\bf 78}, 2803 (1997).

\bibitem{contacts}
If one assumes that the dot is open, $\gamma/\delta\gg{1}$, the
condition for the dynamic localization to be present would be
$\gamma\ll{1}/t_*$, equivalent to $\gamma/\delta\ll\delta/\Gamma$.
Since the theory of Refs.~\cite{us,Basko} is applicable only at
$\Gamma/\delta\gg{1}$, the dot {\em must} be in the Coulomb
blockade regime.

\bibitem{cooldeph}
Note that both for electron escape and phonon emission the ratio
of the cooling rate to the dephasing rate is $\sim{T}^2/\delta$,
i.~e. of the order of the total electronic energy of the dot.
This is due to the fact that each act of escape or phonon emission
leads to a loss of energy of the order of~$T$. As a result, the
relative importance of these mechanisms is the same in cooling
and dephasing.

\bibitem{Bouchiat}
R.~Deblock, Y.~Noat, B.~Reulet, H.~Bouchiat, and D.~Mailly,
Phys. Rev.~B {\bf 65}, 075301 (2002).

\end{thebibliography}
\end{document}